# Sensitive Phase Gratings for X-ray Phase Contrast – a Simulation-based Comparison

Oliver Preusche

Department of Computer Science
Friedrich-Alexander University of Erlangen-Nürnberg (FAU),
Germany
e-mail:  oliver.preusche@fau.de

**Abstract:**
Medical differential phase contrast x-ray imaging (DPCI) promises improved soft-tissue contrast at lower x-ray dose. The dose strongly depends on both the angular sensitivity and on the visibility of a grating-based Talbot-Lau interferometer. Using a conventional x-ray tube, a high sensitivity and a high visibility are somewhat contradicting goals: To increase sensitivity, the grating period has to be reduced and/or the grating distance increased. Technically, this means using a higher Talbot order ($3^{rd}$ or $5^{th}$ one instead of first one). This however reduces the visibility somewhat, because only a smaller part of the tube spectrum will get used. This work proposes to relax this problem by changing the phase grating geometry. This allows to double sensitivity (i.e., double the Talbot order) without reducing the visibility. One proposed grating geometry is an older binary one (75% of a period $\pi$-shifting), but applied in a novel way. The second proposed geometry is a novel one, requiring three height levels for polychromatic correction. The advantage is quantified by a simulation of the resulting interference patterns. Visibilities for the common $\pi$-shifting gratings are compared with the proposed alternative geometries. This is done depending on photon energy and opening ratio of the coherence grating G0. It shows that despite of doubled sensitivity of the proposed gratings, the overall visibility might even improve a little.

## 1. Introduction

In grating-based phase contrast x-ray imaging using a conventional x-ray tube, high sensitivity and high visibility are somewhat conflicting goals. In this work, two new phase grating geometries are proposed, which are better compromises of these two goals in comparison to conventional $\pi$-shifting gratings.

### 1.1 Context

Differential phase contrast measures the change in direction of the wavefront as it passes the sample. Grating-based setups based on the Talbot-Lau interferometer can be applied using conventional x-ray tubes as light source [1], [2]. Talbot-Lau interferometers (Fig. 1) use a phase grating G1 to create a periodic intensity pattern (fringe pattern) at the analyzer grating G2. Each bar of G1 typically shifts the phase by half a wavelength. G2 has to be placed at a specific Talbot distance from G1 (called 1st fractional Talbot order) or at an odd multiple of it ($3^{rd}$, $5^{th}$… fractional Talbot orders). The fringe pattern at G2 behaves exactly like a 'shadow' of

G1 *would* behave. When moving G2 along the x-axis for one period $p_2$, intensities $I$ between $I_{min}$ (i.e., occurs when fringes hit bars) and $I_{max}$ (i.e., when fringes pass slits) get detected (this is called 'phase stepping').

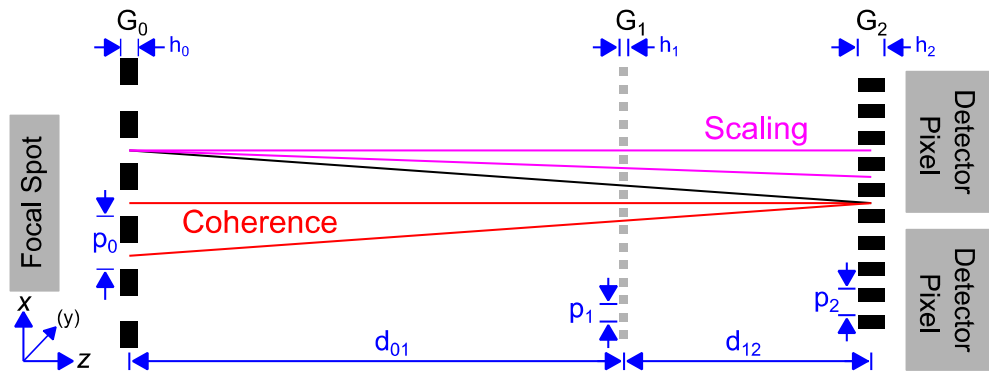

Figure 1.  Talbot-Lau interferometer (shown is a π/2-shifting G1).

## 1.2 Criteria

Using literature [1], [3], [4], it becomes clear that the X-ray dose is proportional to $\alpha^2/V^2$, where $\alpha$ defines the angular sensitivity of the setup and $V$ is the visibility $V = (I_{max} - I_{min})/(I_{min} + I_{max})$ of the setup.

The visibility roughly is the "fringe contrast", i.e., the noise on the measured fringe phase is inversely proportional to it.

The angle $\alpha$ defines the angular sensitivity of the setup (see Fig. 2). The angle $\alpha$ is the factor between fringe phase and the corresponding wavefront direction: Once the fringe phase is determined as an intermediate result, the really interesting wavefront direction can be computed by the fringe phase multiplied by $\alpha/(2\pi)$.

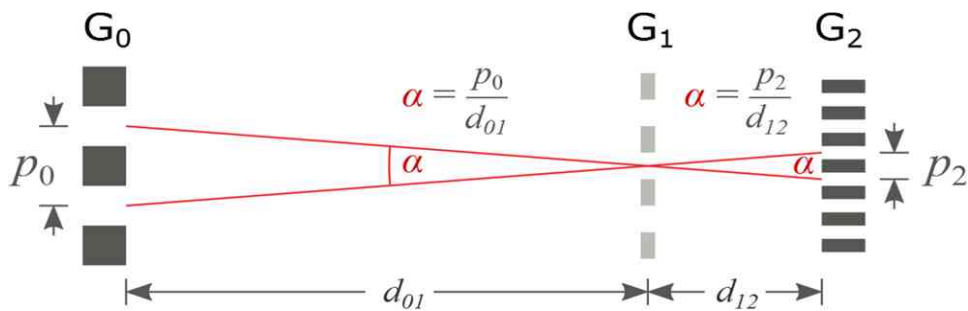

Figure 2.  Angular sensitivity is determined by the angle $\alpha$.

Hence, sensitivity $\alpha$ and visibility $V$ are equally important performance measures. The goal now is to decrease the angle $\alpha$ while keeping the visibility $V$ large.

## 1.3 Fractional Talbot order $m$

A common solution is to choose a higher fractional Talbot order $m$ in the setup. For the basic Talbot-Lau setup, $m=1$ and $d_{12} = p_1^2/(8\lambda) = p_2^2/(2\lambda)$ because $p_2 = p_1/2$ (for simplicity, planar incident wavefronts or equivalently $d_{01} \gg d_{12}$ are assumed here and in the following).

However, the Talbot-Effect also occurs at odd multiples $m$ of the above distance. The quantity $m$ is called fractional Talbot order and the corresponding distance is $d_{12} = m\, p_1^2/(8\lambda)$, so $\alpha_m = \alpha/m$.

The problem is, that the accepted bandwidth of the spectrum decreases when the Talbot order $m$ increases ([1], equation (3)). As an example, [5] (Figs. 4.2 and 4.3) says that $m=5$ has about half the visibility of $m=1$. Fig. 3 below shows that effect ($m=3$ is the graph displayed in blue, $m=1$ is displayed as a gray graph): For the third fractional Talbot order, the visibility is reduced for most photon energies except for the design energy (62 keV).

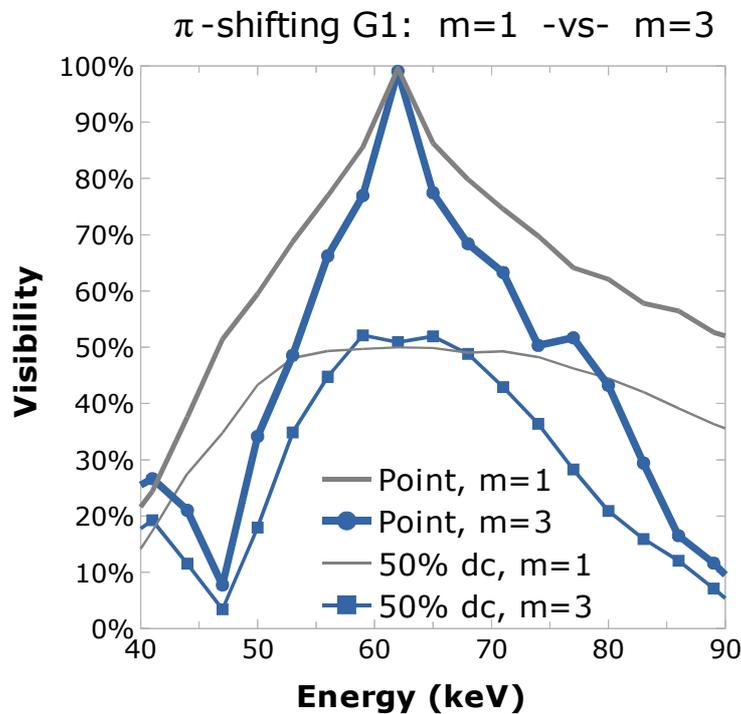

Figure 3. Visibility is shown depending on photon energy for the regular π-shifting G1 grating. The design energy is 62 keV. The gray graphs show m=1, the blue graphs with attached symbols show the third fractional Talbot order m=3. The top pair of graphs is for a point source, the bottom pair for an extended source with source grating (duty cycle 50%).

## 2. Proposal

A long known family of binary gratings ([6], rows 21-28 of table 1) also has a phase shift of $\pi$, but a bar of width $p_1/4$ or (equivalently at design energy) $3/4\ p_1$. The image forms at distance $d_{12}=p_1^2/(4\lambda)$, i.e., at twice the distance for the regular grating, so sensitivity is doubled. The variant with bar width $p_1/4$ has bad polychromatic behavior and is useless. The other one is good and its interference pattern at design energy is shown in Fig. 4a (monochromatic) and Fig. 4b (polychromatic). It consists of a main fringe and a weak secondary fringe. If $p_2 = p_1$ is chosen (i.e., one fringe per $p_1$), then the weak fringe is treated as an unwanted problem. The result is bad sensitivity and visibility. The grating is useful only when choosing $p_2 = p_1/2$ (see the arrow in Fig. 4a). This is shown in Fig. 4c where the polychromatic interference pattern of a second simulated point source superimposes the intensities of Fig. 4b shifted by $p_1/2$ – all fringes are now equivalent. Also, the regions between the fringes stay quite dark, which is important to reach a good visibility.

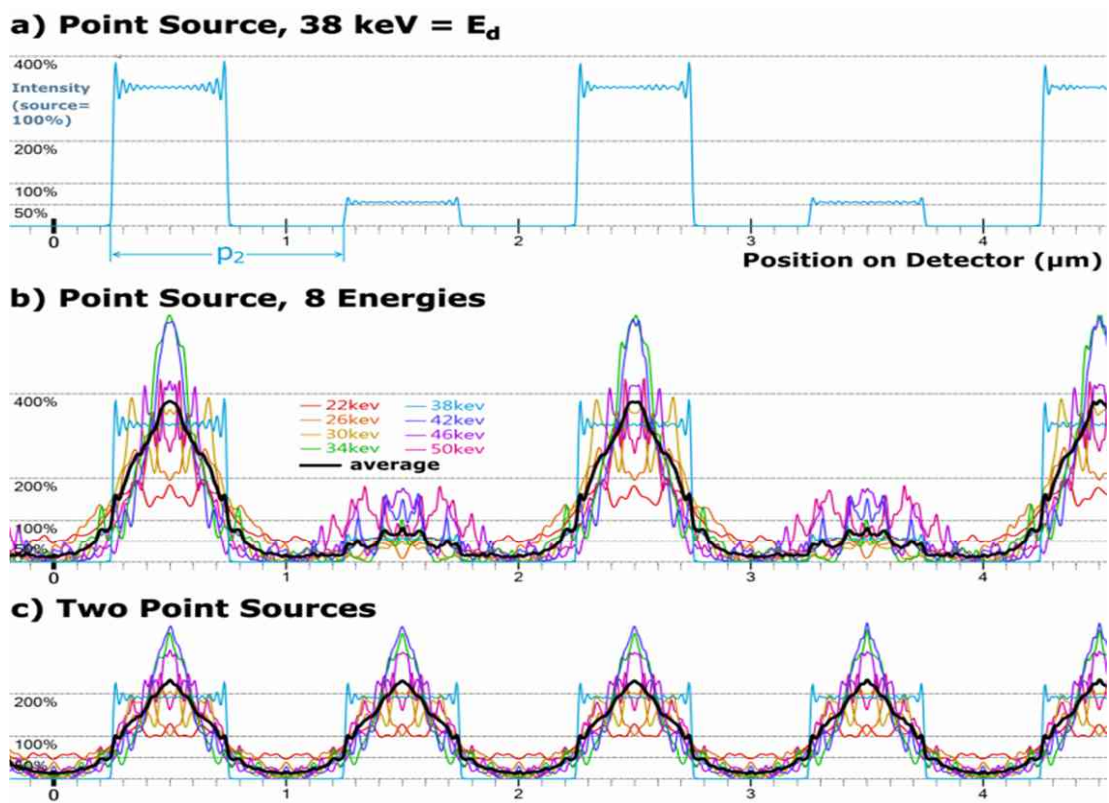

Figure 4. Intensity graph of the first grating ($\pi$-shifting, bar width is $3/4\ p_1$) in G2 plane. $p_1$= 2 µm, $p_2$=1 µm. The "average" graph of the 8 energies is weighted to match a tungsten tube at 55 kVp.

Figure 5 shows resulting simulated visibilities of a nickel-grating of height 13.3 µm, assuming perfectly black G0 and G2 gratings. The design energy is 62 keV. Graphs for a pair of point sources and for a G0 with 50% duty cycle are shown. The dotted lines are the comparison graphs for the regular π-shifting grating. The graph for the point source shows some visibility reduction near the design wavelength, but otherwise the first proposed grating is better than the regular one and furthermore has doubled sensitivity (because of doubled $d_{12}$).

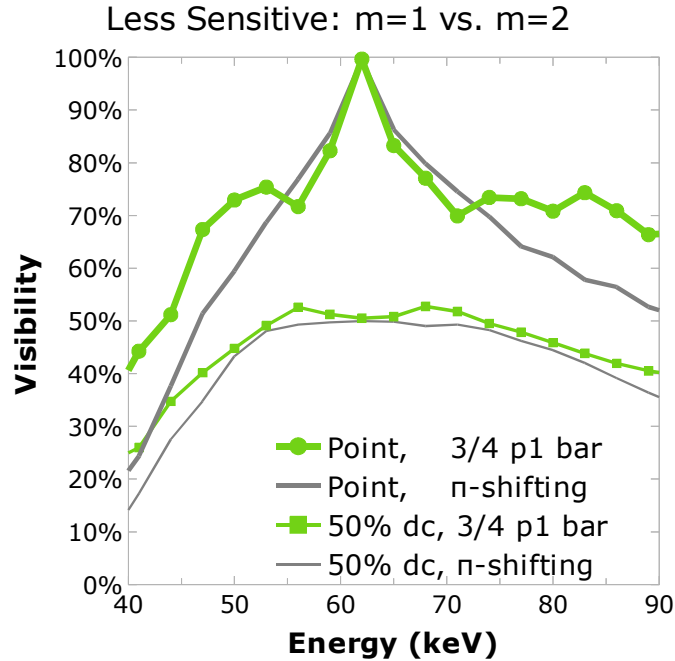

Figure 5. Visibility by photon energy for the first grating ($m = 2$). The gray comparison graphs show the situation for the regular π-shifting grating ($m = 1$). The design energy is 62 keV.

When simulating this first grating at third Talbot Order ($m = 6$), the visibility really drops in case of polychromatic illumination. However, a polychromatic correction is possible: Shifting the phase in the middle $p_1/2$ of the bar by $2\pi$ additionally does not change behavior at design wavelength, but generally increases the visibility as is shown in Fig. 6. It is simulated at distance $d_{12} = 6 p_1^2 / (8\lambda)$, corresponding to $m = 6$ for the regular grating (which has $m = 3$), so it offers doubled sensitivity. As also shown in Fig. 8, this "3-level grating" consists of 4 sections per $p_1$ period of widths 1/4, 1/8, 1/2, 1/8 $p_1$, the corresponding 4 phase shifts are 0, 1/2, 3/2, 1/2 π. Maximum nickel height is 39.9 µm.

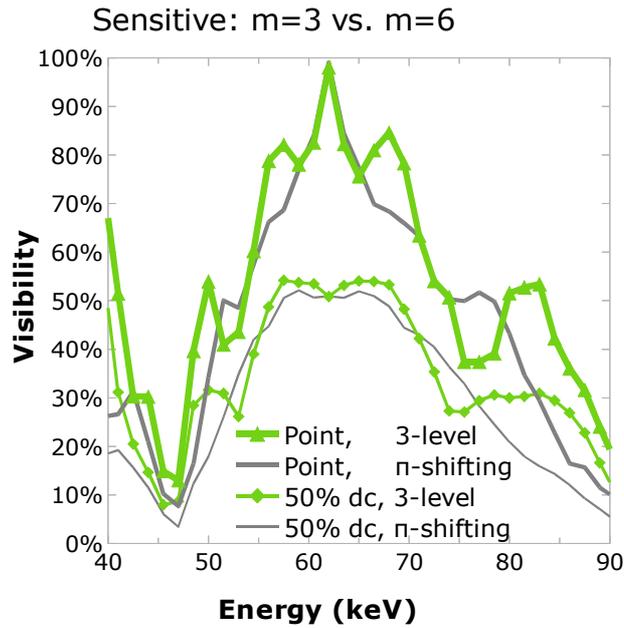

Figure 6. Visibility depending on photon energy for the three-level G1 ($m = 6$). The gray comparison graphs show a π-shifting regular G1 at the $m = 3^{rd}$ fractional Talbot order.

Notice that at third Talbot order, at least 2 further geometries allow high visibilities: Using the section-widths defined above, the phase shifts of the two alternatives are 1/2, 0, 2, 0 π and 0, 1/2, 3/2, 1/2 π.

    The performance of all gratings is summarized in Figure 7. The visibility-graphs of Figs. 3, 4, and 6 were weighted using the spectrum of a Tungsten anode operated at 100 kV, filtered by 200 µm Rhenium (Re) and additionally by 20 µm Gold (Au). Rhenium was used because of its absorption edge at 71.7 keV, a narrow spectrum between 53 and 72 keV results, which matches the visibility graph in Fig. 6. The resulting visibility numbers in Fig. 7 show that the proposed gratings offer higher visibility than the regular π-shifting G1.

    The proposed grating geometries are summarized in Fig. 8. Shown are phase shifts for each section along the x-axis of the grating.

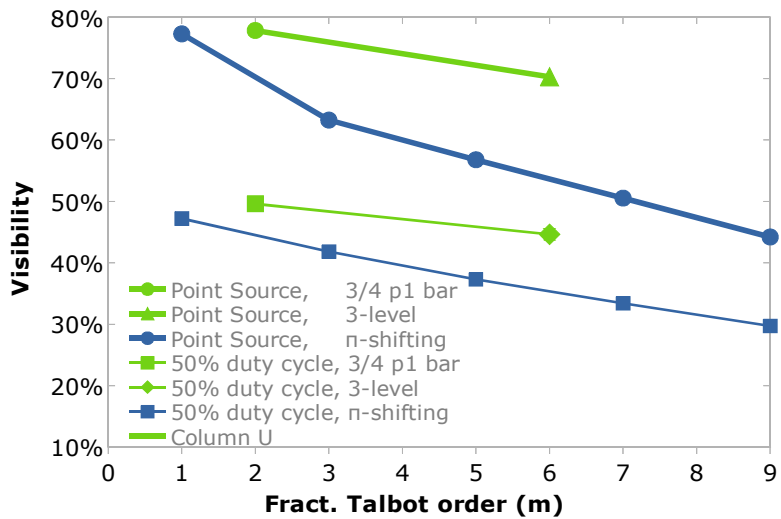

Fig. 7. Summary of visibilities depending on the fractional Talbot order $m$. The blue graphs are for the regular π-shifting gratings, the green graphs show the proposed gratings, Top pair shows point sources, bottom pair extended sources with a source grating of 50% duty cycle.

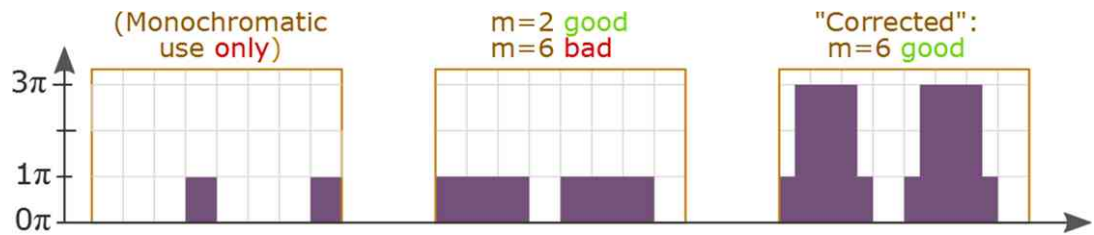

Fig.8. Proposed grating geometries. Each case shows 2 periods of the grating. All three cases differ only in polychromatic light. The left hand side shows a grating which is good only in monochromatic light. The center geometry is the proposed geometry for m=2, the right hand side geometry is the proposal for m=6.

## 3. Conclusion

The first grating is an easy to produce alternative for the first fractional Talbot order with moderately increased absorption. Even at second fractional Talbot order (*m*=2), it might outperform the regular symmetric π-shifting grating at third fractional Talbot order. For the *m*=6$^{th}$ fractional Talbot order, three alternative three-level grating geometries were given. The choice of geometry might depend on grating production technology. They seem to be a good replacement of regular gratings operated at the 5$^{th}$ or 7$^{th}$ Talbot order.

## Acknowledgment

O.P. thanks Prof. Gisela Anton and Thomas Weber of the working group "Phase Contrast" in the Department of Physics for open mindset and discussions.